\documentclass[prc,amsmath,twocolumn,superscriptaddress]{revtex4} 

\usepackage{dcolumn}
\usepackage{graphicx}
\usepackage{color}

\def\beq{\begin{equation}}
\def\eeq{\end{equation}}
\def\bea{\begin{eqnarray}}
\def\eea{\end{eqnarray}}
\newcommand*{\tabref}[1]{Table~\ref{tbl:#1}}
\newcommand*{\tablab}[1]{\label{tbl:#1}}
\renewcommand*{\eqref}[1]{Eq.~(\ref{eq:#1})}
\newcommand*{\eqlab}[1]{\label{eq:#1}}
\newcommand*{\figref}[1]{Fig.~\ref{fig:#1}}
\newcommand*{\figlab}[1]{\label{fig:#1}}
\newcommand*{\appref}[1]{Appendix~\ref{sec:#1}}
\newcommand*{\secref}[1]{Section~\ref{sec:#1}}
\newcommand*{\seclab}[1]{\label{sec:#1}}

\def\VYP#1#2#3{{\bf #1}, #3 (#2)}  
\def\NP#1#2#3{Nucl.~Phys.~\VYP{#1}{#2}{#3}}

\def\NPB#1#2#3{Nucl.~Phys.~B~\VYP{#1}{#2}{#3}}

\def\PLB#1#2#3{Phys.~Lett.~B~\VYP{#1}{#2}{#3}}
\def\PR#1#2#3{Phys.~Rev.~\VYP{#1}{#2}{#3}}

\def\PRD#1#2#3{Phys.~Rev.~D~\VYP{#1}{#2}{#3}}
\def\PRL#1#2#3{Phys.~Rev.~Lett.~\VYP{#1}{#2}{#3}}

\def\ZP#1#2#3{Z.\ Phys.\  \VYP{#1}{#2}{#3}}

\newcommand{\etal}{\mbox{\textit et al.}}                       %

\newcommand{\Omit}[1]{}

\begin{document}

\title{
Optimal Radio Window for the Detection of Ultra-High-Energy Cosmic Rays and
Neutrinos off the Moon.
}

\author{O. Scholten}
\email{scholten@kvi.nl}

\author{J. Bacelar}
\affiliation{Kernfysisch Versneller Instituut, University of Groningen,
9747 AA, Groningen, The Netherlands}

\author{R. Braun}
\affiliation{ASTRON, 7990 AA Dwingeloo, The Netherlands}

\author{A.G. de Bruyn}
\affiliation{ASTRON, 7990 AA Dwingeloo, The Netherlands}
\affiliation{ Kapteyn Institute, University of Groningen, 9747
AA, Groningen, The Netherlands}

\author{H. Falcke}
\affiliation{ASTRON, 7990 AA Dwingeloo, The Netherlands}
\affiliation{Department of Astrophysics,
IMAPP, Radboud University, 6500 GL Nijmegen, The Netherlands}

\author{B. Stappers}
\affiliation{ASTRON, 7990 AA Dwingeloo, The Netherlands}
\affiliation{Astronomical Institute `A. Pannekoek',
University of Amsterdam, 1098 SJ, The Netherlands}

\author{R.G. Strom}
\affiliation{ASTRON, 7990 AA Dwingeloo, The Netherlands}
\affiliation{Astronomical Institute `A. Pannekoek',
University of Amsterdam, 1098 SJ, The Netherlands}

\begin{abstract}
When high-energy cosmic rays impinge on a dense dielectric medium, radio waves
are produced through the Askaryan effect. We show that at wavelengths
comparable to the length of the shower produced by an Ultra-High Energy cosmic
ray or neutrino, radio signals are an extremely efficient way to detect these
particles. Through an example it is shown that this new approach offers, for
the first time, the realistic possibility of measuring UHE neutrino fluxes
below the Waxman-Bahcall limit. It is shown that in only one month of observing
with the upcoming LOFAR radio telescope, cosmic-ray events can be measured
beyond the GZK-limit, at a sensitivity level of two orders of magnitude below
the extrapolated values.
\end{abstract}
\maketitle

\section{Introduction}

The interest in determining the flux of Ultra-High Energy (UHE) cosmic rays and
neutrinos is manyfold. The origin of the highest energy cosmic rays is a major
research topic as the existence of these particles requires very spectacular
events on a cosmic scale. At energies beyond the so-called
Greisen-Zatsepin-Kuzmin (GZK)-limit~\cite{Gre66,Zat66} (at an energy of about
$6\;10^{19}$~eV) the spectrum of cosmic rays is expected to drop rather
drastically. The mechanism for this is that at these energies the cosmic rays
produce pions when scattering off the microwave background while traversing
distances of the order of 10~Mpc. The existence of this cutoff in the spectrum
has not been verified unambiguously up to now. The flux of cosmic rays beyond
the GZK cutoff determines the sources for UHE cosmic rays within a range of
about 10~Mpc~\cite{Ach01,Del04} i.e.\ close on astronomical scales. This is a
very exciting prospect as presently there are no known sources at this
proximity. The presently proposed method is very efficient for determining this
flux.

Another point of interest lies in the detection of UHE neutrinos.  These could
be created by UHE protons producing $\pi^+$ mesons when scattering off the
microwave background (the GZK mechanism as mentioned above) which, through weak
decay, produce neutrinos. These GZK neutrinos have thus far never been
observed.

There are also other, more speculative, models predicting UHE neutrinos. These
models belong to a generic class known as top-down (TD) models, where UHE
particles owe their origin to the decay of some supermassive X-particle of mass
$m_X$. Their decay products, the UHE-cosmic rays, can have energies up to
$m_X$. These massive X particles could be topological defects or magnetic
monopoles that could be produced in the early Universe during symmetry-breaking
phase transitions envisaged in grand unified theories (GUTs);
see~\cite{Yos97,Bha00,Sta04,Sar02} for reviews.

As an efficient method to determine the fluxes of UHE particles we are
investigating the production of radio waves in a particular frequency window
when a UHE particle hits the moon. Askaryan predicted as early as
1962~\cite{Ask62} that particle showers in dense media produce coherent pulses
of microwave \v{C}erenkov radiation. Recently this prediction was confirmed in
experiments at accelerators~\cite{Sal01} and extensive calculations have been
performed on the development of showers in dense media to yield quantitative
predictions for this effect~\cite{Zas92}. The Askaryan mechanism lies at the
basis of several experiments to detect (UHE) neutrinos using the \v{C}erenkov
radiation emitted in ice caps~\cite{Mio04,Leh04}, salt layers~\cite{SALSA}, and
the lunar regolith. The pulses from the latter process are detectable at Earth
with radio telescopes, an idea first proposed by Dagkesamanskii and
Zheleznyk~\cite{Dag89} and later by others~\cite{Alv01}. Several experiments
have since been performed~\cite{Han96,Gor04} to find evidence for UHE
neutrinos. All of these experiments have looked for this coherent radiation
near the frequency where the intensity of the emitted radio waves is expected
to reach its maximum. Since the typical lateral size of a shower is of the
order of 10~cm the peak frequency is of the order of 3~GHz.

Here we propose a different strategy to look for the radio waves at
considerably lower frequencies where the wavelength of the radiation is
comparable in magnitude to typical longitudinal size of showers. It has been
noted before [18] that a new generation of low-frequency digital radio
telescopes will provide excellent detection capabilities for high-energy
particles, thus making our consideration here very timely. We show that the
lower intensity of the emitted radiation, which implies a loss in detection
efficiency, is compensated by the increase in detection efficiency due to the
near isotropic emission of coherent radiation. The net effect is an increased
sensitivity by several orders of magnitude, for the detection of UHE cosmic
rays and neutrinos at frequencies which are one or two orders of magnitude
below that where the intensity reaches its maximum. At lower frequencies the
lunar regolith becomes increasingly transparent for radio waves. This implies
for the detection of UHE neutrinos that there are two gain factors when going
to lower energies; i) Increased transparency of the lunar regolith already
stressed in Ref.~\cite{Fal03}, and ii) Increased angular acceptance, stressed
in this work, which gives much larger count rates.

In the following two sections we explain quantitatively the idea of an optimum
frequency by taking cosmic-ray and neutrino-induced radio-emission from the
Moon as a specific example. The advantage of going to lower frequencies also
applies to other experiments where the radiation crosses a boundary between a
dense medium to one with a considerably lower index of refraction. In Section
IV we propose two specific observations, one for an existing facility, the
Westerbork Synthesis Radio-Telescope array (WSRT), and one for a facility which
will be available in the near future, the Low-Frequency Array (LOFAR).

\section{Model for Radio Emission}

There exist two rather different mechanisms for radio emission from showers
triggered by UHE cosmic rays or neutrinos, where each has received considerable
attention recently. One is the emission of radio waves from a shower in the
terrestial atmosphere. Here the primary mechanism is the synchrotron
acceleration of the electrons and positrons in the shower due to the
geomagnetic field, called geosynchrotron radiation, which has recently been
confirmed with new digital radio
techniques~\cite{Fal03,Sup03,Hue05,Fal05,Ard05}. The second mechanism applies
to showers in dense media, such as ice, salt, and lunar regolith, where the
front end of the shower has a surplus
of electrons. %
Since this cloud of negative charge is moving with a
velocity which exceeds the velocity of light in the medium, \v{C}erenkov
radiation is emitted. For a wavelength of the same order of magnitude as the
typical size of this cloud, which is in the radio-frequency range, coherence
builds up and the intensity of the emitted radiation reaches a maximum. This
process, known as the Askaryan effect~\cite{Ask62} is the subject of this
paper.

The intensity of radio emission (expressed in units of Jansky's where
1~Jy~=~$10^{-26}$~W~m$^{-2}$Hz$^{-1}$) from a hadronic shower, with energy
$E_s$, in the lunar regolith, in a bandwidth $\Delta\nu$ at a frequency $\nu$
and an angle $\theta$, can be parameterized as (see \appref{AngSpr})
\bea
&&\hspace*{-1.5em} F(\theta,\nu,E_s)
 = 3.86 \times 10^4\; e^{-Z^2} \Big( {\sin{\theta}\over \sin{\theta_c}} \Big)^2
 \Big( {E_s \over 10^{20} \mbox{ eV} } \Big) ^2
 \nonumber \\ &&\hspace*{-1.5em}  \times
 \Big( {d_{moon} \over d } \Big)^2
 \Big( {\nu \over \nu_0 (1+(\nu/\nu_0)^{1.44})} \Big)^2
 ({\Delta\nu \over 100\mbox{ MHz}}) \; \mbox{Jy} \;,
 \eqlab{shower-l}
\eea
with
\beq
Z
 = (\cos{\theta} -1/n)
 \Big({n\over \sqrt{n^2-1}}\Big)\Big({180\over \pi \Delta_c}\Big)\;,
 \eqlab{Z}
\eeq
where $\nu_0=2.5$~GHz~\cite{Gor04}, $d$ is the distance to the observer, and
$d_{moon}=3.844 \times 10^8$~m is the average Earth-Moon distance. The angle at
which the intensity of the radiation reaches a maximum, the \v{C}erenkov angle,
is related to the index of refraction ($n$) of the medium,
$\cos{\theta_c}=1/n$. Crucial for our present discussion is the spreading of
the radiated intensity around the
\v{C}erenkov angle, given by $\Delta_c$ (in degrees). The $\sin{\theta}$ factor in
\eqref{shower-l} reflects the projection of the velocity of the charges in
the shower on the polarization direction of the emitted \v{C}erenkov radiation.
The dependence of $Z$ as defined in \eqref{Z} is suggested by working out some
specific cases~\cite{Leh04}, see \appref{AngSpr}. For small values of
$\Delta_c$ it coincides with the formula \eqref{shower} found in much of the
literature~\cite{Zas92,Alv01,Gor04} however, \eqref{shower-l} is more accurate
for large spreading angles.

The spreading of the radiated intensity around the \v{C}erenkov angle,
$\Delta_c$, is, on the basis of general physical arguments, inversely
proportional to the shower length and the frequency of the emitted radiation.
Based on the results given in Ref.~\cite{Alv01} it can be parameterized as
\beq
\Delta_c = 4.32^\circ
 \Big({1\over \nu\,[\mbox{GHz}] }\Big)
 \Big({L(10^{20}\mbox{eV})\over L(E_s) }\Big) \;,
\eqlab{del_c_had'}
\eeq
where $L(E_s)$ is the shower length which depends on the energy. In
Ref.~\cite{Alv98}, calculated results are given for the shower length (in units
of radiation lengths, equal to 22.1~g/cm$^{2}$~\cite{Alb73} for lunar regolith)
which at the highest energies can be parameterized as
\beq
L(x)=12.7 + {2\over 3} x \;,
\eqlab{Le}
\eeq
where $x=\log_{10}(E/10^{20}\mbox{ eV})$. At an energy of $10^{20}$~eV this
corresponds to a shower length of approximately 1.7~m, where we take the
density of the regolith to be approximately 1.7~g/cm$^3$. For a frequency
$\nu=200$~MHz, where the wavelength is of the same order as the shower length,
we should expect on general arguments that the radiation spreads over an
angular range ($2\Delta_c$) that is comparable to the \v{C}erenkov angle,
$\theta_c=56^\circ$. This indeed corresponds to the value $\Delta_c =
21.52^\circ$, obtained from \eqref{del_c_had'}. In \appref{AngSpr} the
predictions for the angular spread based on the parametrization
\eqref{shower-l} using
\eqref{del_c_had'} is compared with analytical calculations for some
(simplified) shower profiles showing excellent consistency with the calculated
shower length.

In our simulations we have taken into account the attenuation of radio waves in
the regolith. As a mean value for the attenuation length for the radiated power
we have taken $\lambda_r= (9/\nu$[GHz])~m~\cite{Olh75}, which is the same value
as used in the analysis of the GLUE experiment~\cite{Gor04}. This value is
obtained from loss-tangent measurements performed on samples of lunar basalt
brought back from the Moon~\cite{Olh75}. The measured values show a rather
large variation, the effects of which are investigated in
\secref{Accuracy}. In principle the layer of regolith is only 10-20 m thick
under which there a thick layer of fractured rock going over into solid
bedrock. As shown in \appref{LunReg} the bedrock is as efficient in emitting
radio waves at the lower frequencies as the regolith. At higher frequencies,
due to the larger attenuation length, only the relatively thin upper layer
contributes to radio emission. All in all this implies that for the calculation
of the acceptance the structure of the deeper layers (rock v.s.\ regolith) is
not important. In the calculations we have therefore included radiation coming
from a depth of at most 500 m treating for simplicity, and without loss of
accuracy, the whole layer as behaving like regolith. It is argued in
\appref{LunReg} that this indeed gives a realistic estimate for the
acceptance calculations.

In the calculations for cosmic-ray-induced showers we assumed that the shower
occurs effectively at the lunar surface. As argued in \appref{ShaSho} only a
very small depth is necessary for \v{C}erenkov radiation to be emitted. For
neutrino-induced showers an energy-dependent mean free path~\cite{Gan00} has
been used, $\lambda_\nu= 130\; \Big( {10^{20} \mbox{ eV} \over E_\nu }
\Big)^{1/3}$~km which is appropriate for regolith.

A crucial point in the simulation is the refraction of radio waves at the lunar
surface as was already stressed in Ref.~\cite{Zas92}. Since the index of
refraction of the lunar regolith is relatively large, $n=1.8$ corresponding to
a \v{C}erenkov angle of $\theta_c=56^\circ$, much of the radio wave which is
emitted by the shower suffers internal reflection at the surface. Only
radiation with an angle of $90^\circ-\theta_c$ or less with respect to the
normal to the surface will be emitted from the Moon. Since most showers, being
cosmic ray or neutrino induced, are directed towards the center of the Moon,
internal reflection will severely diminish the emitted radiation at high
frequencies where the
\v{C}erenkov cone is rather narrow. The major advantage of going to lower
frequencies is that the spreading $\Delta_c$ around $\theta_c$ increases,
allowing for the radiation to escape from the lunar surface. With decreasing
frequency the peak intensity of the emitted radiation decreases (see
\eqref{shower}), however the peak intensity increases with increased
particle energy. The net effect is that at sufficiently high shower energies
the aforementioned effect of increased spreading is far more important,
resulting in a strong increase in the detection probability.

\begin{figure}
    \includegraphics[height=7.9cm,bb=17 112 742 627,clip]{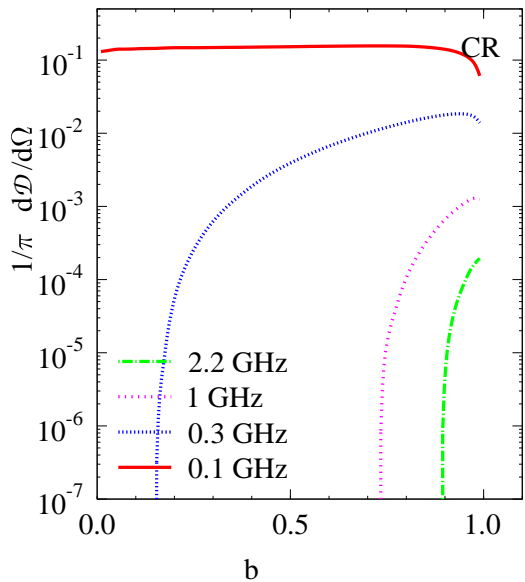}
\caption[fig1]{Differential detection probability, d${\cal D}/$d$\Omega$, for a
cosmic ray of energy $4 \times 10^{21}$ eV hitting the Moon as function of
apparent distance from the center of the Moon, $b$, for different detection
frequencies.}
  \figlab{NM_b-h}
\end{figure}

To be able to address these issues quantitatively we introduce the detection
efficiency ${\cal D}(E,\nu)$. It is defined as the probability that a cosmic
ray (or neutrino) hitting the Moon (at an arbitrary angle and position) with
energy $E$ would produce radio waves at frequency $\nu$ which is detectable at
Earth. We regard an event to be detectable when the power of the signal is 25
times larger than the ``noise level'' which we take equal to
$F_{noise}=20$~Jy~\cite{noise}, i.e.\ a minimal detection level of 500~Jy is
taken using a bandwidth of $\Delta\nu=20$~MHz in \eqref{shower-l}. These values
can be considered typical for LOFAR, see \secref{RealObs}. It is illustrative
to plot the differential detection probability, d${\cal D}/$d$\Omega$, for a
particular area d$A= R^2 $d$\Omega$ on the Moon where $R$ is the lunar radius.
In
\figref{NM_b-h}, d${\cal D}/$d$\Omega$ is plotted for cosmic rays at an energy
of $4 \times 10^{21}$~eV for different frequencies as a function of the
relative distance from the center of the face of the Moon, $b$, where $b=1$
corresponds to the rim of the Moon. At lower frequencies the length of the
shower becomes comparable to the wavelength of the radiation. Coherent emission
of radio waves thus happens over a large angular range instead of only within a
narrow cone at the
\v{C}erenkov angle. The effect of this is that even cosmic rays hitting the
center of the face of the Moon (as seen by us) at very oblique angles are
detectable. If the radiation were emitted in a very narrow cone around the
\v{C}erenkov angle, much of the radiation, even if the cone is directed towards
the Earth, would be be internally reflected off the lunar surface. All this
implies that at high frequencies only the rim of the Moon contributes to
d${\cal D}/$d$\Omega$ and that this contribution is not really large since only
very inclined cosmic rays may produce detectable emission. At lower frequencies
the whole surface of the Moon contributes with a relatively large probability
since a large range of angles contribute. This trend is clearly discernable in
\figref{NM_b-h}.

\begin{table}
  \caption{ \tablab{Luminos}
Calculated detection probabilities for cosmic-ray induced showers at
$E_{cr}=4\times 10^{21}$~eV, labelled as ${\cal D}_{cr}(40)$, and for
neutrino-induced showers at $E_\nu=2\times 10^{22}$~eV, labelled as ${\cal
D}_\nu(200)$, for different frequencies $\nu$ (in units of GHz). }
  \begin{tabular}{c|c|c}
 $ \nu $ & $ {\cal D}_{cr}(40)  $ & $ {\cal D}_\nu(200) $ \\
    \hline
 $ 2.2 $ & $ 1.92 \times 10^{-5} $ & $ 3.03 \times 10^{-7} $ \\
 $ 1.0 $ & $ 2.65 \times 10^{-4} $ & $ 4.85 \times 10^{-6} $ \\
 $ 0.3 $ & $ 9.84 \times 10^{-3} $ & $ 1.97 \times 10^{-4} $ \\
 $ 0.1 $ & $ 1.44 \times 10^{-1} $ & $ 2.92 \times 10^{-3} $ \\
  \end{tabular}
\end{table}

In \tabref{Luminos} the detection probabilities, ${\cal D}$ (d${\cal
D}/$d$\Omega$ integrated over the lunar surface), are given as a function of
frequency. This shows a strong increase of ${\cal D}$ with decreasing
frequency. For a given shower the radiation that is transmitted through the
surface (which is not internally reflected) is proportional to $\Delta_c^2$,
where the quadratic dependence is due to the fact that the phase space is in
both the polar and in the azimuthal angle. An additional factor of $\Delta_c$
comes from the fact that also for showers making an angle of up to $\Delta_c$
with respect to the tangent to the surface, radio waves may be transmitted
through the surface. In total one expects thus ${\cal D}\propto\Delta_c^3
\propto \nu^{-3}$ which agrees rather well with the numbers given in \tabref{Luminos}.

For neutrino-induced showers only 20\% of the initial energy is converted to a
hadronic shower, while the remaining 80\% is carried off by the lepton. This
energetic lepton will not induce a detectable radio shower. For a muon, the
density of charged particles will be too small, while the shower of a UHE
electron will be extremely elongated due to the Landau-Pomeranchuk-Migdal
effect~\cite{Alv98}. The width of the \v{C}erenkov cone will thus be very small
which makes the shower practically undetectable. For the present calculations
we therefore have limited ourselves to the hadronic part of the shower which
carries 20\% of the energy of the original neutrino. To be able to compare the
results for cosmic-ray and neutrino-induced showers in
\tabref{Luminos} the latter have been calculated for a 5-times higher energy of
the incoming particle.

While the showers for hadronic cosmic rays develop very close to the lunar
surface, those of UHE neutrinos are distributed over a rather large range of
depths from the surface. As a result of the long attenuation length for
neutrinos in matter (tens of km at our energies), a large fraction of the
neutrinos create showers too deep inside the Moon so that the radio waves are
attenuated to below the detectable threshold at Earth. Roughly, one thus
expects that the ${\cal D}$ for neutrino-induced showers will be a factor
$\lambda_r/\lambda_\nu$ smaller than ${\cal D}$ for cosmic-ray-induced showers
at the same shower energy. This factor explains much of the difference between
${\cal D}_{cr}$ and ${\cal D}_\nu$ given in
\tabref{Luminos}.

A deviation from the simple $\lambda_r/\lambda_\nu$ scaling occurs since, due
to the relatively long neutrino mean free path, the hadronic part of the
neutrino-induced shower has a certain probability of being directed towards the
surface (in contrast to showers induced by hadronic cosmic rays which are
always directed into the Moon). This increases the probability that part of the
radio signal is emitted from the Moon. The latter can be seen from
\figref{NM_b-n} where at 2.2~GHz the acceptance ring is broader than in
\figref{NM_b-h}. 

\begin{figure}
    \includegraphics[height=7.9cm,bb=17 112 742 627,clip]{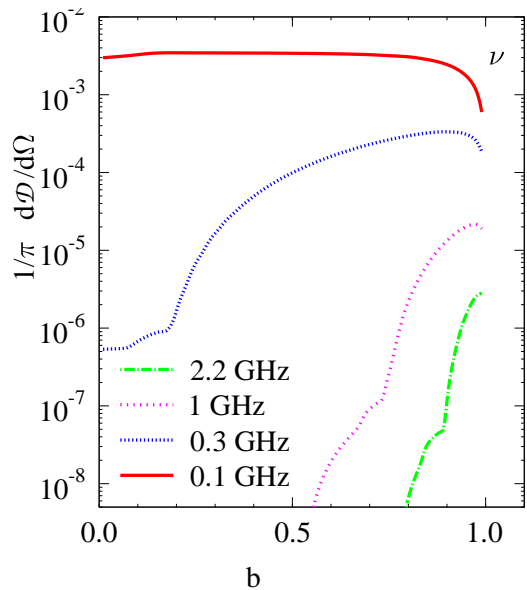}
\caption[fig2]{Same as \figref{NM_b-h} but for showers induced by a neutrino
of energy $2 \times 10^{22}$~eV. It is assumed that 20\% of the neutrino energy
is deposited in an hadronic shower, the shower induced by the high-energy
lepton is ignored in the present calculation since its \v{C}erenkov cone is
very sharp and thus a negligible fraction of its energy will penetrate through
the lunar surface.}
  \figlab{NM_b-n}
\end{figure}

An additional advantage of using lower frequencies is that the sensitivity of
the model simulations to large- or small-scale surface roughness is diminished.
Since at lower frequencies already a sizable fraction of the radiation
penetrates the surface, its roughness will not make a major difference. This is
in contrast to high frequencies where most of the radiation is internally
reflected when surface roughness is ignored.

The increased spreading of the radiation around the \v{C}erenkov cone at lower
frequencies strongly increases the detection efficiency but at the same time
decreases the sensitivity to the direction of the original cosmic ray or
neutrino. Part of this can be recovered by measuring the polarization direction
of the radio waves. The electric field is 100\% polarized in the direction of
the shower, which has been confirmed in laboratory experiments~\cite{Sal01}.

\section{Detection Limits}

\begin{figure}
    \includegraphics[height=7.9cm,bb=36 137 515 672,clip]{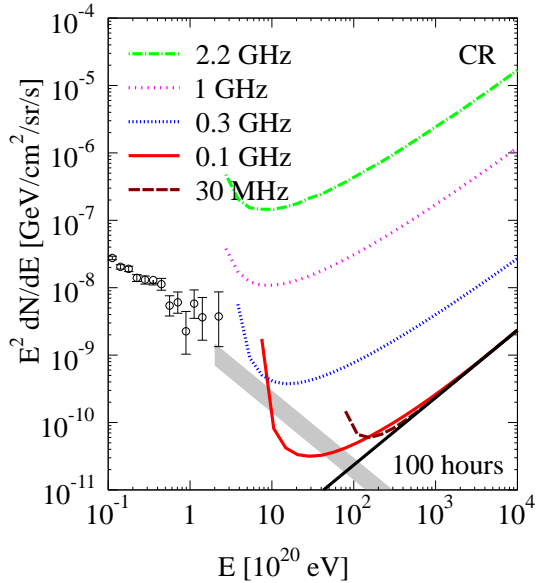}
\caption[fig3]{Flux limits (assuming a null observation) for cosmic rays as can be determined in a
100 hour observation (see text). In the curves for $\nu=30$~MHz a ten fold
higher detection threshold is used, corresponding to the higher sky temperature
at this frequency. The points given correspond to the data of the AGASA
experiment~\cite{Tak03}, the grey band is an extrapolation of the HiRes
data~\cite{Abb04}. The thick black line corresponds to the best possible limit
(vanishing detection threshold). }
  \figlab{Flux-h}
\end{figure}

In \figref{Flux-h} the detection limits for UHE cosmic rays for different
radio-frequency ranges are compared with data from the AGASA~\cite{Tak03}
(points) and a linear extrapolation based on the data from the HiRes
experiment~\cite{Abb04} (grey bar). The model-independent limit is defined as
$dN/dE_{lim}=Q/({\cal D}\times h)$ where $h$ equals the observation time in
hours and $Q =1.16 \times 10^{-22}$ cm$^{-2}$s$^{-1}$sr$^{-1}$ is the full
phase-space for 1 hour on the Moon, assuming an isotropic distribution for the
cosmic rays or neutrinos and assuming the whole face of the Moon is in the
antenna field of view. As in the previous section we have calculated the
detection probability ${\cal D}$ for a signal threshold of 500~Jy at all
frequencies. It should be noted that for $\nu=30$~MHz there is a strong
increase in the sky temperature and we have used a ten-fold higher threshold.
As a result of the higher detection threshold the flux limit lies considerably
higher.

From \figref{Flux-h} one clearly sees that with decreasing frequency one loses
sensitivity for lower-energy particles. This follows directly from
\eqref{shower-l} since with decreasing frequency the maximum signal strength
decreases and thus one exceeds the detection threshold only for more energetic
particles. If the energy of the cosmic ray is more than a factor 4 above this
threshold value the analysis presented in the previous section applies and the
detection limit improves rapidly with decreasing radio-frequency until one
reaches a frequency of 100~MHz where one obtains the optimum sensitivity.
Decreasing the frequency even lower provides no gain since the detection limit
has already reached the optimum $dN/dE_{opt}=Q/(0.5\times h)$, given by the
heavy black line in \figref{Flux-h}.

\begin{figure}
    \includegraphics[height=7.9cm,bb=36 137 515 672,clip]{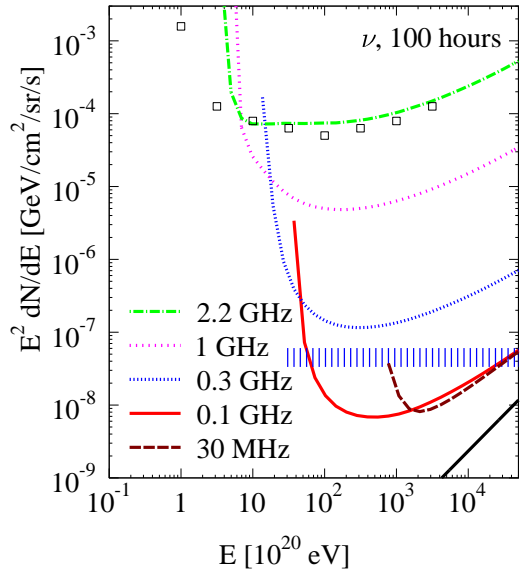}
\caption[fig4]{Similar flux limits as shown in \figref{Flux-h} but for UHE
neutrinos. The open squares are the limits determined from the GLUE
experiment~\cite{Gor04}.}
  \figlab{Flux-n}
\end{figure}

In \figref{Flux-n} we compare the detection limits for UHE neutrinos at
different frequencies with the results obtained from the GLUE
experiment~\cite{Gor04}. One sees similar trends as in the predictions for
cosmic rays, in particular the large gain in the determined flux limits with
decreasing frequency. At higher energies the limits for neutrinos do not
increase as steeply as those for cosmic rays. This is because the neutrino
mean-free-path decreases with energy, therefore increasing the probability for
the neutrino to initiate a shower close to the surface where the attenuation of
the radio waves is small. Our result at 2.2~GHz happens to lie close to that of
the GLUE experiment. The dashed-dotted curve in \figref{Flux-n} shows the
results of a calculation where we have reproduced the simulation for the GLUE
experiment, i.e.\ included (in a somewhat simplified manner) the effects of
averaging over lunar-surface slopes of 10$^\circ$, including a 10\% coverage of
the Moon, and have used the appropriate detection threshold. This result lies
close to the published limits of the GLUE experiment.

In this work we mainly address the detection limits for fluxes of UHE cosmic
rays or neutrinos and a distinction between the two does not have to be made.
When one would measure one or more events the question of distinguishing
between the two kinds becomes interesting. For a single event the frequency
dependence of the pulse (assuming a broad band acceptance) might give an
indication since for deep showers the high-frequency part will suffer a larger
attenuation than the low-frequency part. For a large number of events the
distribution over the lunar surface could be exploited.

\section{WSRT \& LOFAR predictions\seclab{RealObs}}

The Westerbork Synthesis Radio Telescope (WSRT) consists of fourteen 25~m
parabolic dishes located on an east-west baseline extending over
2.7~km~\cite{WSRT}. It is normally used for super-synthesis mapping, but
elements of the array can also be coherently added to provide a response
equivalent to that of a single 94~m dish. Observing can be done in frequency
bands which range from about 115 to 8600~MHz, with bandwidths of up to 160~MHz.
The low frequency band which concerns us here covers 115-170~MHz~\cite{WSRT-l}.
Each WSRT element has two receivers with orthogonal dipoles enabling
measurement of all four Stokes parameters. In tied-array mode the system noise
at low frequencies is $F_{noise}=$600~Jy. To observe radio bursts of short
duration, the new pulsar backend (PuMa II) will be used. It can provide
dual-polarization baseband sampling of eight 20~MHz bands, enabling a maximum
time resolution of approximately the inverse of the bandwidth. In the
configuration which we propose to use, four frequency bands will observe the
same part of the moon with the remaining four a different section. In total,
coverage of about 50\% of the lunar disk can be achieved.

An even more powerful telescope to study radio flashes from the moon will be
the LOFAR array~\cite{LOFAR}. With a collecting area of about 0.05~km$^2$ in
the core (which can cover the full moon with an array of beams), LOFAR will
have a sensitivity about 25 times better than that of the WSRT.  LOFAR will
operate in the frequency bands from 30-80 and 115-240~MHz where it will have a
sensitivity of about $F_{noise}=$600~Jy and $F_{noise}=$20~Jy, respectively.
The Galactic background noise will become the dominant source of thermal noise
fluctuations at frequencies below about 100~MHz.  It therefore appears that the
optimal radio window for the detection of cosmic ray or neutrino induced radio
flashes from the moon will be around 100-150~MHz.

\begin{figure} 
    \includegraphics[height=7.9cm,bb=27 137 515 672,clip]{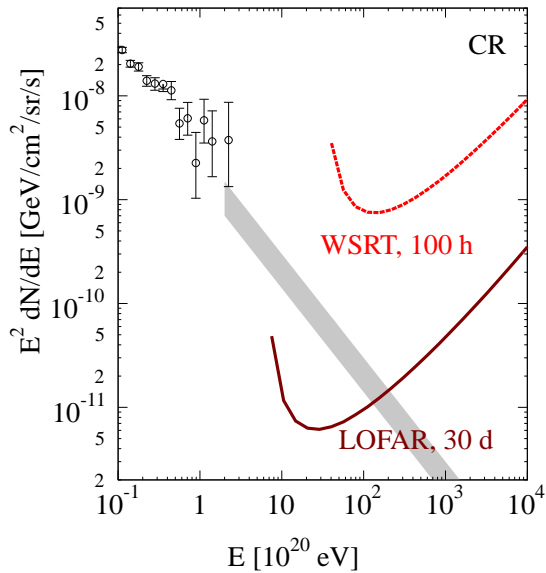}
\caption[fig5]{Flux limits on UHE cosmic rays as can be determined
in a 30 day observation with the LOFAR antenna system and a 100 hour
observation with WSRT. The data shown are the same as in
\figref{Flux-h}.}
  \figlab{LOFAR-h}
\end{figure}

The probability for a observing a signal simultaneously in four frequency bands
with a power of 20$\times F_{noise}$ per frequency band, where $F_{noise}$ is
the noise level per polarization direction, as a random fluctuation of the
background noise is less than 0.001 (equal to $3 \sigma$ significance) for a
100 h observing period at a sampling rate of 40 MHz, typical for our PUMA-II
back-end. Simulations show that a pulse of intensity 25$\times F_{noise}$,
interfering with the noise background, can be detected with $3 \sigma$
significance at a probability greater than 80\%. For this reason we have
assumed in the calculations a detection threshold of 25$\times F_{noise}$ for
both the WRST and the LOFAR telescopes.

A simulation for LOFAR, taking $\nu=120$~MHz, bandwidth of $\Delta\nu=20$~MHz,
a signal-detection threshold of 500~Jy,  and an observation time of 30 days is
shown in \figref{LOFAR-h} for cosmic rays and in
\figref{LOFAR} for neutrinos. The results are compared with the limits that can
be obtained from a presently proposed observation for 100 hours at the WSRT
observatory assuming a detection threshold of 15,000~Jy, $\nu=140$~MHz,
bandwidth of $\Delta\nu=20$~MHz, and a 50\% Moon coverage.

In \figref{LOFAR} the predicted spectrum of GZK neutrinos is taken from
Ref.~\cite{Eng01}. The prediction of top-down (TD) neutrinos is based on the
decay of a topological defect with a mass of $10^{24}$~eV as was calculated in
Ref.~\cite{Sta04,Pro96}. The Waxman-Bahcall (WB) limit~\cite{Bah01}, based on
theoretical arguments, is an upper limit on the neutrino flux which is
consistent with the data on the fluxes of UHE cosmic rays. The limits are also
compared with those from the GLUE~\cite{Gor04} and FORTE~\cite{Leh04}
experiments which are calculated as model independent limits similar to our
limits. The limit from the RICE~\cite{Kra03} experiment has been calculated
assuming different power-law spectra for the neutrinos. In general such a limit
lies below the model-independent limit~\cite{Leh04}.

With the existing WSRT a limit on the neutrino flux can be set which falls just
above the WB bound. However, even this will constrain different top-down
scenarios, discussed in the literature. With the proposed LOFAR facility this
limit can be improved considerably to reach, for the first time, a limit well
below the WB bound for neutrinos. In addition one has a good chance to see
evidence (in only a 30 day period) of cosmic ray events at an energy one order
of magnitude higher than presently observed.

\begin{figure}
    \includegraphics[height=7.9cm,bb=27 137 515 672,clip]{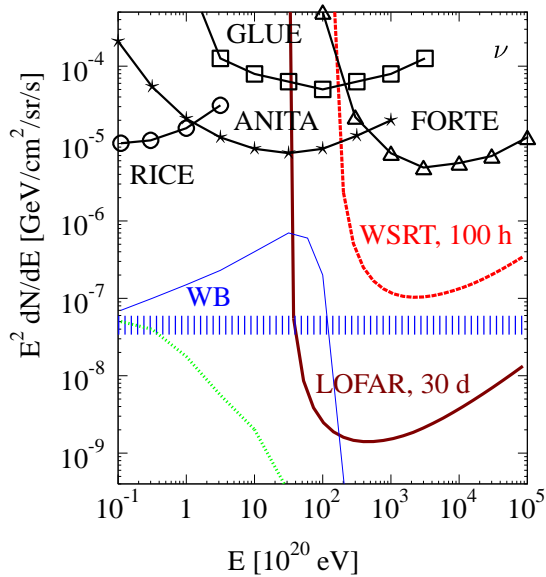}
\caption[fig6]{Flux limits on UHE neutrinos as can be determined
with WSRT and LOFAR observations (see text) are compared with various models,
in particular, WB~\cite{Bah01} (vertical bars), GZK~\cite{Eng01} (dotted thin
line), and TD~\cite{Sta04,Pro96} (solid thin line). Limits from the
RICE~\cite{Kra03}, GLUE~\cite{Gor04}, ANITA~\cite{Bar06}, and
FORTE~\cite{Leh04} experiments are also shown.}
  \figlab{LOFAR}
\end{figure}

\section{Accuracy of the Predictions \seclab{Accuracy}}

\begin{figure} 
    \includegraphics[height=7.9cm,bb=27 137 515 672,clip]{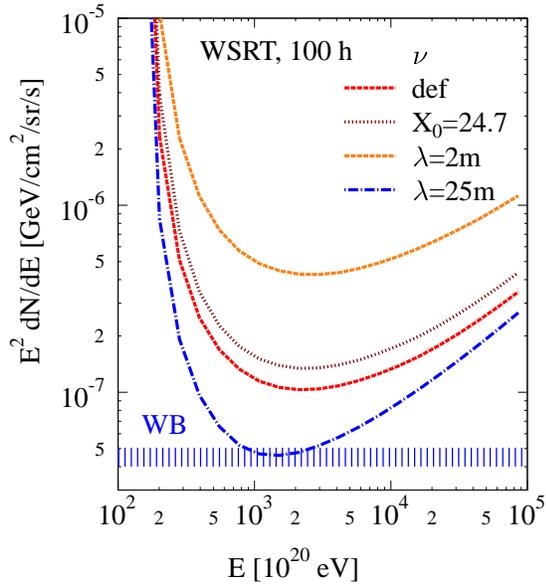}
\caption[fig7]{Model dependence of flux limits on UHE neutrinos for WSRT.
Note the different scale in this figure.}
  \figlab{Errors-nu}
\end{figure}

In this section we address the robustness of our acceptance calculations. Some
of the sources of model-dependence in the calculation have already been
discussed. A potentially large one is due to the uncertainty in the thickness
of the regolith layer. As is argued in \appref{LunReg}, the final result is
rather insensitive to this thickness since the rock below the regolith is as
efficient, if not more so, in emitting radio-waves at the frequencies we are
interested in.

The loss tangent of the regolith, which determines the attenuation of radio
waves, is found to be dependent on the metallic composition of the regolith.
Different Apollo samples show a large variation. The extremes for
radio-attenuation distance in regolith vary between 2/$\nu$ m/GHz to about
25/$\nu$ m/GHz~\cite{Olh75}. We have used an intermediate value of 9/$\nu$
m/GHz for our estimates. As can be seen from \figref{Errors-nu} (please note
the expanded scale) the two extremes for the radio-absorbtion distance result
in roughly a variation in the acceptance for neutrinos which is equal to the
variation in the radio-attenuation length. The reason is that the thickness of
the layer of the lunar crust which can be `seen' on earth is proportional to
the attenuation length of radio waves. It should be realized that the showers
initiated by hadronic cosmic rays are close to the surface for which the exact
value of the loss-tangent is unimportant.

\begin{figure} 
    \includegraphics[height=7.9cm,bb=27 137 515 672,clip]{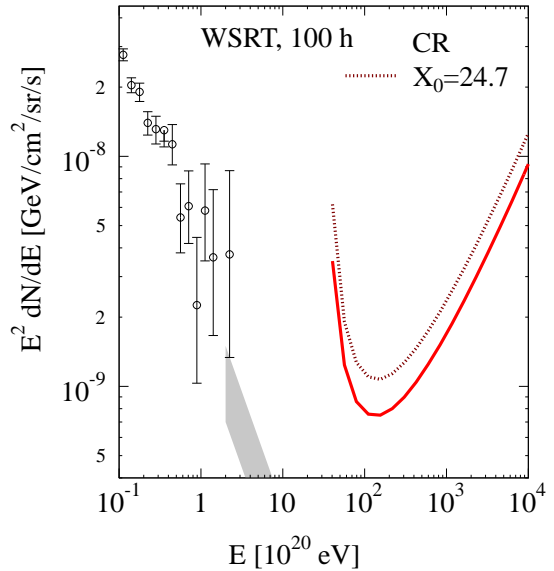}
\caption[fig8]{Model dependence of flux limits on UHE cosmic rays for WSRT
(and the LOFAR antenna system).}
  \figlab{Errors-h}
\end{figure}

We show in \appref{AngSpr} that the angular spread of the emitted radio-waves
from a shower depends on the length of the shower. This length is proportional
to the stopping power $X_0$. As the stopping power is mostly determined by
electronic processes it is strongly dependent on the elements in the rock or
regolith. We have determined the stopping power for some of the Apollo samples
analyzed in Ref.~\cite{Alb73,Mey04}. This shows that the actual variation in
the stopping power is not as large as one might have expected. We have
determined $X_0=22.1$~g/cm$^2$ for the `A-17 HIGH Ti'~\cite{Alb73} sample which
was used in the Monte Carlo simulations~\cite{Alv98}. It is a typical Lunar
Basalt and is similar to samples 70017, 70035 in ref.~\cite{Mey04}. For the
`A-15 Pyroxene'~\cite{Alb73} sample we determined $X_0=23.2$~g/cm$^2$ and
$X_0=24.7$~g/cm$^2$ for some Apollo-16 samples~\cite{Kor97}. In order to obtain
an estimate of the sensitivity of the acceptance to the stopping power we also
show in
\figref{Errors-nu} the results of a calculation where the angular spread is
reduced by a factor, equal to the reduction of the shower length, (24.7/22.1).
Since the detection limits are proportional to the third power of the angular
spread, this results in only a 40\% change.

An additional source of model dependence lies in principle in the details of
modelling the roughness of the lunar surface. For the GLUE experiment, at much
shorter wavelength, this roughness gave rise to a considerable broadening of
the angular acceptance and thus to a large increase in the acceptance. Since,
for the wavelength of interest for the LOFAR and WRST telescopes, the angular
spread is already large, the additional effects of surface roughness can be
ignored as they give rise to only a minor increase in the detection
probability.

The Puma-II back-end which is available at the WRST telescope allows for
storing all data on disk for each observation period. A total of 12 hours of
observing time can be stored. The analysis of the data can thus be done
off-line allowing full flexibility in optimizing the corrections for
ionospheric dispersion. Because of the Nyquist sampling of the signal in the
PUMA-II back-end the pulse may cover -after correction for dispersion- two to
three sampling times. In the off-line analysis this can be taken into account.
In the LOFAR operation the signal is also stored but only after a trigger
condition is met in a fast on-line signal analysis.

\section{Summary}

We have demonstrated the clear advantage of using radio waves at frequencies
well below the \v{C}erenkov maximum. The optimum frequency will be that where
the length of the shower, of the order of several meters in the lunar regolith,
is of the same order of magnitude as the wavelength of the radio waves where
the radio-emission pattern is nearly isotropic. The advantage of going to lower
frequencies applies to all experiments where the radiation crosses a boundary
between a dense medium to one with a considerably lower index of refraction.

We have shown that the gain in efficiency at lower frequencies is such that
with the upcoming LOFAR facility one can seriously investigate realistic
top-down scenarios for UHE neutrinos and be sensitive to neutrino fluxes well
below the Waxman-Bahcall limit. Even now with the existing WSRT, profiting from
its capability to measure right in the radio-frequency window where the
detection efficiency is highest, one is able to set limits on neutrino fluxes
orders of magnitude below the present limit in only a 100~h observation period.

For UHE cosmic rays the LOFAR facility offers, because of the availability of
an optimal radio-frequency window, a very powerful tool to determine the flux
beyond the GZK limit. In only a 30~day observing period one is sensitive to a
flux which is more than one order of magnitude below the extrapolation of the
measured flux from below the GZK limit. Since the beam of LOFAR is determined
by software, much longer observation periods should also be attainable,
resulting in a sensitivity to even lower fluxes. Assuming the GZK limit is
real, this offers the exciting possibility to measure the density of sources
for UHE cosmic rays within a range of about 10~Mpc.

As an additional topic one may access the composition of cosmic-rays (proton
versus heavy nuclei) by searching for the predicted coulomb dissociation of
heavy nuclei passing in the neighborhood of the high density photon field of
the Sun, the so called Gerasimova-Zatsepin effect~\cite{Ger60}. The original
predictions were recently revised~\cite{Med99}, showing that the separation of
the two dissociated daughter nuclei at 1 AU from the Sun is as large as
hundreds or even thousands of km, making the moon an excellent detector. The
difference in the time of arrival of the two particles determines the mass of
the original cosmic-ray nucleus.

\begin{acknowledgments}
This work was performed as part of the research programs of the Stichting voor
Fundamenteel Onderzoek der Materie (FOM) and of ASTRON, both with financial
support from the Nederlandse Organisatie voor Wetenschappelijk Onderzoek (NWO).
We gratefully acknowledge discussions with J.~Alvarez-Mu\~{n}iz on different
aspects of shower development in dense media.
\end{acknowledgments}

\appendix

\section{Angular Spread \seclab{AngSpr}}

Since the angular spread of the intensity of the \v{C}erenkov radiation around
the \v{C}erenkov angle for the case $\lambda\approx L$ is crucial for our
considerations, we will present here a discussion of this case which is
independent of the parameterizations given in the literature at shorter
wavelength.

In the literature the intensity of \v{C}erenkov radiation from a hadronic
shower, with energy $E_s$, in the lunar regolith, in a bandwidth $\Delta\nu$ at
a frequency $\nu$ and an angle $\theta$, has been parameterized, based on Monte
Carlo simulations, as~\cite{Zas92,Alv01,Gor04}
\bea
F(\theta,\nu,E_s)&=& 3.86 \times 10^4\; e^{-((\theta-\theta_c)/\Delta_c )^2}
 \Big( {d_{moon} \over d } \Big)^2  \nonumber \\ &&
 \times  \Big( {E_s \over 10^{20} \mbox{ eV} } \Big) ^2
 \Big( {\nu \over \nu_0 (1+(\nu/\nu_0)^{1.44})} \Big)^2 \nonumber  \\ &&
 \times ({\Delta\nu \over 100\mbox{ MHz}}) \; \mbox{Jy} \;,
 \eqlab{shower}
\eea
where $\nu_0=2.5$~GHz and the spreading $\Delta_c$ is given by
\eqref{del_c_had'}. \eqref{shower} has been experimentally shown~\cite{Sal01}
to be accurate within a factor 2 for a shower of particles that would
correspond to a primary particle of about $10^{19}$~eV at frequencies exceeding
1 GHz where the wavelength is small compared to the longitudinal extent
(length) of the shower. For the case in which the wavelength is comparable to
the length of the shower, of interest for the present investigation, one may
expect deviations from the simple parametrization where our main concern is the
angular spread of the \v{C}erenkov radiation.

To focus on the angular spread, we have derived the angular distributions for
two different shower profiles following the approach given in~\cite{Leh04}. For
the first one, called the ``block'' profile, the number of charged particles is
constant over the shower length $L=L_b$, $\rho_b(x)=1$ for $0<x<L_b$. This
profile is not realistic for a shower as the full intensity suddenly appears
and disappears. For this reason we have also investigated a second profile
where the charge in the shower appears and disappears following a sine profile,
$\rho_s(x)=\sin{\pi x/L_s}$ with $0<x<L_s$.

\begin{figure}
    \includegraphics[height=7.9cm,bb=50 147 515 650,clip]{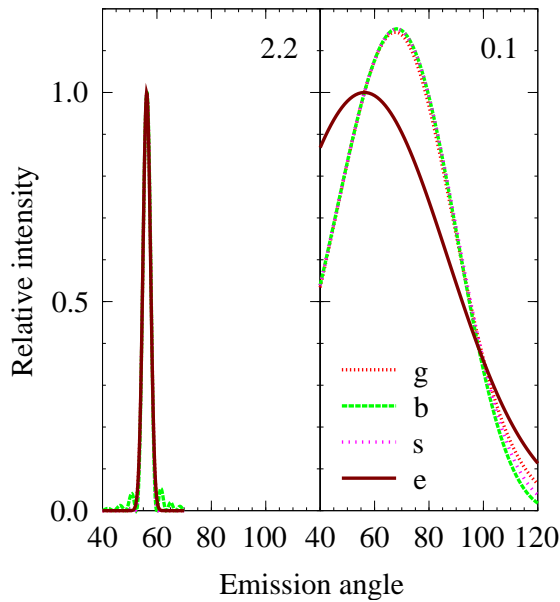}
\caption[fig9]{The angular spread around the \v{C}erenkov angle for the block
and the sine shower-profile functions are compared to the parametrization used
in this work (e: \eqref{shower}; g: \eqref{I-gaus}; b: \eqref{I-block}; s:
\eqref{I-sin}). The left (right) hand displays the results for 2.2 GHz
(100MHz) respectively. }
  \figlab{ang-spread}
\end{figure}

For the block longitudinal profile we reproduce the well known intensity
distribution found by Tamm~\cite{Tam39} for a finite length shower, normalized
to unity at the \v{C}erenkov angle,
\beq
I_b(\theta) = \Big[ {\sin{\theta}\over \sin{\theta_c}}
 {\sin{\pi \chi} \over \pi \chi} \Big]^2
\eqlab{I-block}
\eeq
with
\beq
\chi=(\cos{\theta} -1/n)L/\lambda
\eeq
For the normalized ``sine'' profile we obtain
\beq
I_s(\theta) = \Big[ {\sin{\theta}\over \sin{\theta_c}}
 {\cos{\pi \chi}\over (1-2\chi)(1+2\chi)} \Big]^2
\eqlab{I-sin}
\eeq
The predictions of these two formulas are compared with the parametrization of
\eqref{shower} at 2.2~GHz and 100~MHz for a shower of $10^{20}$~eV.
To reproduce the angular spread of this calculation at 2.2~GHz we choose
$L_b=2.5$~m and $L_s=L_b \times 4/3=3.4$~m, see \figref{ang-spread}. The
results are also compared to those of the gaussian parametrization proposed
in~\cite{Leh04},
\beq
I_s(\theta) = \Big[ {\sin{\theta}\over \sin{\theta_c}} \Big]^2
 e^{-Z^2} \;,
\eqlab{I-gaus}
\eeq
with $Z$ given by \eqref{Z-App}. The value for $Z_0$, \eqref{Z0-App}, is chosen
such that for a small angle expansion around $\theta_c$ it agrees exactly with
\eqref{shower}.
From the figure it is seen that the simple parametrization of \eqref{shower} is
reproduced well by all three analytic forms. \eqref{I-block} shows the well
known secondary interference maxima, due to the sharp edges of the
profile~\cite{Bun01}, which are not realistic for our case. Keeping parameters
fixed the angular distributions are now compared at 100 MHz (right hand panel
of \figref{ang-spread}). The three analytic forms, Equations
(\ref{eq:I-block}), (\ref{eq:I-sin}), and (\ref{eq:I-gaus}) agree quite
accurately but differ considerably from
\eqref{shower}. The reason for this difference lies mainly in the pre-factor
$\sin^2{\theta}$, missing in \eqref{shower}, which accounts for the radiation
being polarized parallel to the shower and thus that emission at 0$^\circ$ and
180$^\circ$ is not possible.

On the basis of the arguments given above we will use the gaussian
parametrization, accurate at small and large angles,
\bea
&&\hspace*{-1em}F(\theta,\nu,E_s) = 3.86 \times 10^4\; e^{-Z^2} \Big(
{\sin{\theta}\over
\sin{\theta_c}} \Big)^2
 \Big( {d_{moon} \over d } \Big)^2
  \\ && \times
 \Big( {E_s \over 10^{20} \mbox{ eV} } \Big) ^2
 \Big( {\nu \over \nu_0 (1+(\nu/\nu_0)^{1.44})} \Big)^2
 ({\Delta\nu \over 100\mbox{ MHz}}) \; \mbox{Jy} \;, \nonumber
 \eqlab{shower-App}
\eea
with
\beq
Z=(\cos{\theta} -1/n)Z_0 \;.
 \eqlab{Z-App}
\eeq
The value for
\beq Z_0=\Big({n\over \sqrt{n^2-1}}\Big)\Big({180\over \pi \Delta_c}\Big)
\eqlab{Z0-App} \;,
\eeq
whth $\Delta_c$ measured in degrees (see \eqref{del_c_had'}. For small
spreading angles around the \v{C}erenkov angle, i.e.\ short wavelengths, this
expression agrees with the results of Monte Carlo
simulations~\cite{Zas92,Alv01,Gor04} while at large wave lengths the formula
agrees with the analytic results~\cite{Alv05}.

As a last point we compare the length of the shower, 1.7 m according to
\eqref{Le}, with the value for the length used in \eqref{I-sin}, $L_s=3.4$ m.
It should be realized that for the sine profile, only for half its length
(i.e.\ 1.7 m) the density of charged particles exceeds 70\% of the maximum
value, which is the definition of the shower length in the Monte-Carlo
simulations. The agreement is thus excellent.

\section{Shallow Showers \seclab{ShaSho}}

In this paper we have implicitly assumed that the showers develop well inside
the lunar regolith. For the emission of long wavelength radio waves from cosmic
ray showers which are close to the lunar surface this assumption needs a more
detailed consideration. The proximity of the surface, through mechanisms like
mirror charges, could severely diminish the amount of \v{C}erenkov radiation
through the surface.

Even though the general problem has not been studied, the -in some sense-
inverse problem has been studied~\cite{Ulr66}, namely that of an electron beam
in close proximity to a dielectric. When the velocity of the electrons exceeds
that of the velocity of light in the medium ($c_m=c/n$, where $n$ is the index
of refraction) \v{C}erenkov radiation is induced in the medium. The occurrence
of this process has been verified experimentally~\cite{Tak00}. In
Ref.\cite{Ulr66} the amount of \v{C}erenkov radiation is calculated as a
function of all key parameters in the problem such as the distance $a$ of the
electron beam from the surface, the electron velocity $\beta=v/c$, the
wavelength of the radiation $\lambda$, and the angle $\eta$ of the radiation
with respect to the surface of the dielectric. In the limit of
ultra-relativistic particles ($\beta=1$) the dependence of the intensity on $a$
reads
\beq
W \sim e^{-4 \pi \, a \sqrt{n^2-1}/\lambda} \;.
\eqlab{Ulr}
\eeq
This equation differs from that quoted in~\cite{Tak00} where instead the limit
$\beta\gtrsim 1/n$ has been used. \eqref{Ulr} clearly shows that proximity of
the surface is only an issue when ${4 \pi \, a \sqrt{n^2-1}/\lambda}\lesssim 1$
or for $a \lesssim \lambda/(4 \pi \sqrt{n^2-1})=0.05 \lambda$ for the lunar
regolith. This implies that this effect should be considered only for showers
making very small angles to the surface, i.e.\ ``shallow showers'' with
$sin{\theta}\lesssim 0.05 \lambda /(0.5 \, L)$, where $L$ is the shower length.
For $L=\lambda$ this corresponds to an angle of less than $6^\circ$.

To test the contribution of shallow showers to our results we have made
calculations in which the contribution of shallow showers are excluded. The
difference in the results is barely visible in the plots and this effect can
thus safely be ignored.

It should however be realized that it is only an extrapolation to apply the
conclusions based on the work of Ref.\cite{Ulr66} to the present problem. A
thorough theoretical treatment should be performed.

\section{Lunar Regolith \seclab{LunReg}}

The properties of the regolith play an important role in the present
calculations. The index of refraction and the loss tangent (determining the
attenuation length of radio waves) have been determined from samples brought
back to Earth in the Apollo missions. From the average values one extracts an
index of refraction of $n=1.8$ and an attenuation length of $\lambda_r=
(9/\nu$[GHz])~m~\cite{Olh75,Hei91} for the intensity of radio waves in
regolith.

\begin{figure}
    \includegraphics[height=7.9cm,bb=36 142 510 807,clip]{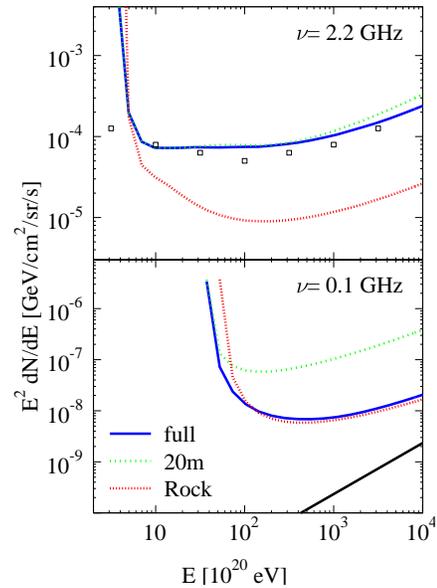}
\caption[fig10]{The dependence of the detection limits on properties of the
lunar regolith. The solid curve represents a calculation made as if the
regolith extends to a depth of 500 m, with only 20 m assumed for the dotted
curve, while pure lunar rock is taken for the short dashed calculation.}
  \figlab{regol}
\end{figure}

One issue of particular interest for the detection efficiency of
neutrino-induced showers is the thickness of the regolith. In his
thesis~\cite{Tak03th} Takahashi made an extensive study of the absorbtion and
reflection of radio waves in the frequency domain of 0.1 to 10 MHz for
realistic depth profiles based on~\cite{Hei91}. He expects a smoothly varying
attenuation length of $\lambda_r= 4.08 (\nu$[GHz]$)^{-.81}$~m for depths
ranging from 200 m up to 100 km. At these depths one expects a decreased
attenuation length due to the increased density of the material. Since the
density of rock is about twice as high as that of the regolith also the value
of the index of refraction (and thus the \v{C}erenkov angle) and the length of
hadronic shower (and thus the spread around the \v{C}erenkov angle) should be
modified. To study these different effects we compare in \figref{regol} the
results of three different calculations. In the first (the solid curve) we
assume that the properties of the regolith layer apply to depths of 500 m. In
the second (dotted curve) the regolith expends to a depth of 20 m and we assume
that no radiation from deeper layers reaches the surface. In the third
calculation we assume that the lunar rock extends to the surface of the moon.
The density of the rock is taken twice that of the regolith, the index of
refraction equal to 2.6, the radio absorbtion length equal to one third of that
of the regolith, which takes into account the larger value of the loss tangent
in rock.

At a frequency of 2.2 GHz one finds, by comparing the drawn and the dashed
curves, that only showers in the upper part of the regolith are detectable due
to the relatively strong absorbtion of radio waves. At this frequency the
detectability of showers in rock is much higher due to the fact that the spread
around the \v{C}erenkov angle is twice as large due to the reduced length of
the shower which in turn is a consequence of the larger density. However the
rock is at most areas covered by a layer of regolith which will completely
attenuate the radio waves. The calculations for 20 or 500 m of regolith (hardly
any difference between the two) thus yield a conservative lower limit. At
100~MHz the situation is quite different. The calculations for pure rock or
pure regolith give rise to very similar limits. The increase in spreading width
for the calculation in rock is apparently compensated by an increased
attenuation. Due to the larger wavelength the contribution of showers deep
under the surface are important. If only the showers in the upper 20 m of
regolith are taken into account in the calculation, the limit changes by about
1 order of magnitude. This calculation completely ignores the emission from
deeper showers which is clearly unrealistic. If one adds the contribution from
the deeper rock layer one will obtain a result close to the drawn curve.
Including the fact that for rock covered by a layer of regolith the index of
refraction changes more gradually, giving rise to reduced reflection near the
surface, would even give a more stringent lower limit. Again, the calculation
taking a maximum depth of 500 m gives a realistic estimate for the limit.

The calculations in the main part of this work account for radiation emitted
from a depth of not more that 500 m. On the basis of the arguments presented in
the above it will be clear that this gives a realistic estimate for lower
bounds on neutrino fluxes. For cosmic rays none of these considerations are
important as all induced showers lie in the upper part of the regolith.

\end{document}